\documentclass[twocolumn,showpacs,preprintnumbers,amsmath,amssymb]{revtex4}

\usepackage{graphicx}
\usepackage{dcolumn}
\usepackage{bm}

\begin{document}

\preprint{CaN2006}

\title{New ferromagnetic nitrides, CaN and SrN, and their synthesis process}
\author{Masaaki Geshi, Koichi Kusakabe, Hitose Nagara, and Naoshi Suzuki}
\affiliation{%
Department of Engineering Science, Osaka University\\
1-3 Machikaneyama, Toyonaka, Osaka, Japan
}%

\date{\today}

\begin{abstract}
We introduce new type of ferromagnets, CaN and SrN, which were designed
using first-principles calculations. 
These are half-metallic ferromagnets and they have magnetic moments of 1 
$\mu_{\rm B}$ per chemical formula unit.
Out of the typical structures of binary compounds, the rock-salt
 structure is the most stable form for both CaN and SrN.   
The majority of the magnetic moment of these compound originates from
 the N sites since the $p$ states of N are spin-polarized.
Their formation energies were calculated and the results show that it
 should be feasible to synthesize these materials. 
The structural stability of CaN was confirmed by performing
 first-principles molecular dynamics simulations. 
We propose a synthesis process for CaN basd on the first-principles.
\end{abstract}


\maketitle

Synthesis of substances which do not exist in nature promises to have an  
enormous potential to produce materials having excellent properties.  
Recent experimental reports of the syntheses of PtN,\cite{PtN}
IrN, IrN$_2$ and OsN$_2$\cite{IrN} demonstrate that it is 
possible to synthesize new materials using special experimental
techniques.  

On the other hand, the reliability of standard first-principles
calculations has been widely recognized. 
They are very powerful tools for predicting the physical and chemical
properties of materials including the individual characteristics of
elements based on their electronic configurations. 
In particular, structural properties and physical quantities calculated
from them highly accurate are in good agreement with experimentally
measured values. 
To date there has been no example of an incorrect prediction of the
magnetic properties of simple transition metals compounds using a
spin-dependent generalized gradient approximation (spin-GGA) for an
exchange-correlation functional.    
These facts imply that it is possible to obtain accurate predictions of  
structural and magnetic properties without performing experiments.  
This allows us to design new materials before they are synthesized in
the laboratory.  
Our aim is to design new materials and discover the synthesis processes
of these new materials using theoretical techniques.  

In this study, we focus on nitride compounds.
It is well known that nitrides prefer the ferromagnetic state. 
For example, most rare earth nitrides become ferromagnetic\cite{RE} and
usually crystallize in a rock-salt(RS) structure. 
These facts encourage us to consider replacing the rare earth elements
with Ca and Sr since their ionic radii are close to those of the rare
earth ions, and from our previous results we confirmed that Ca- and
Sr-pnictides are ferromagnetic\cite{Geshi_ICPS-27}.  
Such replacement is natural in view of the Zintl phase\cite{Zintl}. 
There is thought to a probability of being able to synthesize RS-CaN and 
RS-SrN if appropriate experimental techniques
are employed. 
In fact, PtN has been synthesized in a high pressure
environment\cite{PtN}.

Many predictions have been made\cite{Shirai, Sato,CrO2, Mazin} based on 
density functional theory (DFT)\cite{HK} using the local spin density
approximation (LSDA) or the spin-GGA, and these predictions have been
experimentally
confirmed.\cite{Akinaga,GaMnN,GaMnN_Sasaki,CrO2_E,CrO2_Soulen,FeCoS2}   
Spin-GGA calculations of the electronic structures and the magnetic
properties of materials which consist of atoms which have only $s$ and
$p$ electrons tend to be the most reliable, and those of second-row
elements such as carbon and oxygen, give satisfactory results.  

We performed first-principles calculations based on DFT using the LSDA
or the spin-GGA for RS-CaN and RS-SrN.   
Typical DFT codes were used to confirm our results
\cite{Wien2k,Machikaneyama2002,Espresso}. 
The total energies of some structures and each magnetic state and
detailed electronic structures (e.g. densities of states (DOS) and
magnetic moments) were calculated by the APW+lo method.\cite{Wien2k}  
The value of $R_{MT} K_{max}$ was fixed at 8.00, where $R_{MT}$ is the
minimum muffin-tin (MT) radius and $K_{max}$ is the maximum reciprocal
lattice vector.
The MT radii of Ca and Sr (MT$_{\rm Ca(Sr)}$)are 0.22$a$, 0.27$a$ and   
0.45$a$ for the zinc-blende(ZB), RS, and CsCl structures, respectively,
where $a$ is the lattice constant.  
The MT radii of N (MT$_{\rm N}$) are 0.18$a$, 0.221$a$, and 0.3681$a$
for the ZB, RS, and CsCl structures, respectively.
For the wurtzite structure MT$_{\rm Ca(Sr)}$ and MT$_{\rm N}$ were set
to fixed values for each volume and ratio between MT$_{\rm Ca(Sr)}$
and MT$_{\rm N}$ is 1.3:1 and the sum of both MT radii is about 95\% of
atomic distances. 
We used an angular momentum expansion up to $l_{max}=10$.
The energy convergence criterion was set to 0.001 mRy.
Twenty k-points were taken in the irreducible Brillouin zone.
To confirm magnetism in the four types of structures, we performed more
accurate calculations which employed ten times larger k-point sampling
in the irreducible zone.  
By doing this we were able to confirm the definite existence of
magnetism for these cases. 
To check the magnetism, we performed calculations for RS-CaN
using the KKR-CPA-LDA method \cite{Machikaneyama2002}.
For Morruzi-Janak-Williams method was used to calculate the
exchange-correlation functional \cite{MJW}.
For this calculation we used an angular momentum expansion up to
$l_{max}=2$. 
We used the Espresso-code\cite{Espresso} to perform first-principles 
molecular dynamics simulations and also to check the magnetism of 
RS-CaN .  
The energy cut-off of plane waves was 80 Ry.
$8 \times 8 \times 8$ k-point meshes were used for first Brillouin zone 
integration.
Other details concerning the calculations are described later.


\begin{figure}
\includegraphics[width=8.0cm]{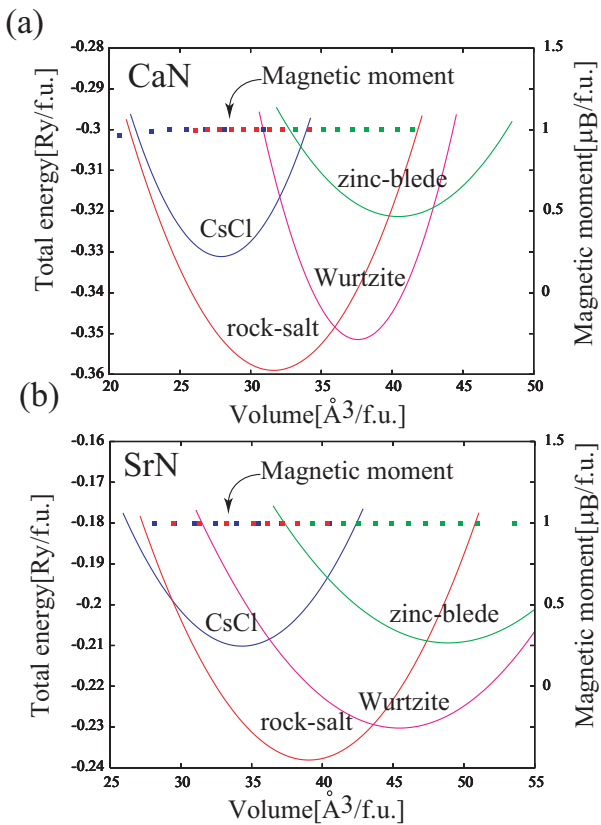}
\caption{\label{fig:1} (Color online)
Total energies and magnetic moments of (a) CaN and (b) SrN are plotted 
as a function of volume. 
The magnetic moment per chemical formula unit is shown on the 
right vertical axis. 
}
\end{figure}
\begin{figure}[t]
\includegraphics[width=7.0cm]{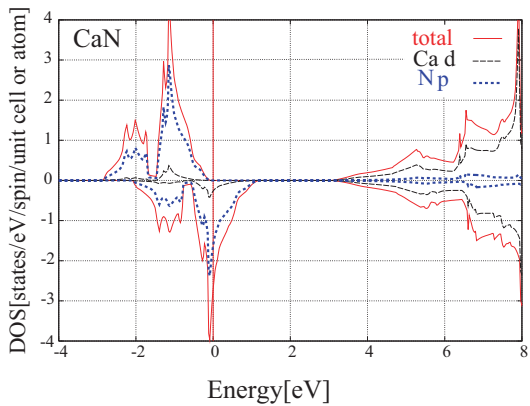}
\caption{\label{fig:2} (Color online)
The DOSs of RS-CaN. 
The partial DOS is defined in the muffin-tin sphere used in the band
 calculations. 
}
\end{figure}

In Figs. \ref{fig:1}(a) and \ref{fig:1}(b), the total energies of CaN
and SrN are plotted as a function of volume per chemical formula unit.  
We chose the RS structure along with the other typical structures 
of binary compounds, namely, ZB, wurtzite and CsCl structures.   
The most lowest structure is the RS structure, while the wurtzite
structure is the second lowest stable structure. 
The wurtzite structure was optimized with respect to both $c/a$ and the
internal parameter.  
We checked the stabilities against tetragonal distortion for the RS, ZB
and CsCl structures and confirmed that the RS and the CsCl structures
are stable.  
In particular, the stability of the cubic structure for the RS
structure is doubly confirmed on the basis of these total energy
calculations and the first-principles molecular dynamics simulations
described below. 
By contrast, tetragonal distortion reduces the total energy of the
ZB structure, which is similar to the case of CaAs in the ZB
structure\cite{Geshi_ICPS-27}.   
In the ZB structure with tetragonal distortion, c/a=1.0 is the
maximum energy point and c/a=0.6 is the minimum energy point.  
The energy difference between these two points is about 0.53eV(0.039Ry)
for CaN and 0.23eV (0.017Ry)for SrN.  
Based on these differences, the distorted ZB structure is
thought to have almost the same energy as the wurtzite one 
(Figs. \ref{fig:1}(a),(b)).

The four structures used in the calculations for CaN and SrN are of
half-metal as well as ferromagnetic in spite of not containing any
transition or rare earth metals.   
In other words, CaN and SrN are ferromagnetic without having any $d$ or
$f$ electrons. 
The magnetic moments are plotted in Figs. \ref{fig:1}(a) and (b). 
The calculated magnetic moments for both CaN and SrN
are 1 $\mu_{\rm B}$, which is evidence of their half-metallicity.
The DOSs of RS-CaN are shown in Fig. \ref{fig:2}. 
The total DOS indicates completely spin-polarized electronic states at
the Fermi level.    
In the other structures, CaN and SrN are also half-metallic as shown
in Fig. \ref{fig:1}, while the magnetic moments of Ca(P,As,Sb) and
Sr(P,As,Sb) depend on the structures and lattice constants or
volumes.\cite{Geshi_ICPS-27} 
The partial DOSs indicate that the principal component at the Fermi
level consists of $p$ states of N. 
As expected, the majority of the magnetic moment resides on the N atom
as shown by Table \ref{table_CaN}.
The magnetic moment at the Ca or Sr sites is quite small being less than
that of the interstital region.   

\begin{table}[tb]
\caption{
The magnetic moment of CaN. In units of $\mu_{\rm B}$.
\label{table_CaN}  }
\begin{ruledtabular}
\begin{tabular}{lccc}
structure & RS & ZB & CsCl \\
\hline
Toatl     & 1.000  &  1.000 & 1.000 \\
Ca        & 0.038  &  0.038 & 0.033 \\
N         & 0.794  &  0.787 & 0.822 \\
Interstitail & 0.168 & 0.175 & 0.145 
\end{tabular}
\end{ruledtabular}
\end{table}

\begin{figure}[tb]
\includegraphics[width=7.5cm]{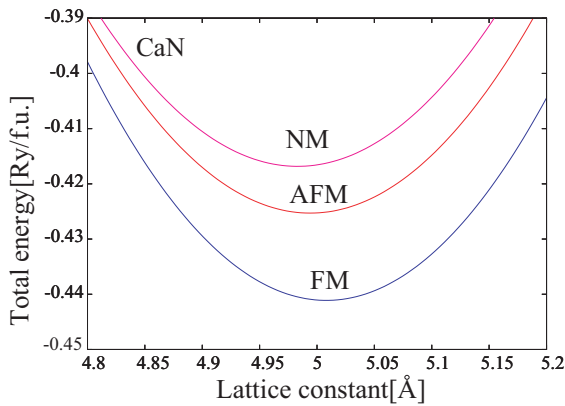}
\caption{\label{fig:3} (Color online)
Total energies as a function of the lattice constant are compared 
for the ferromagnetic (FM), antiferromagnetic (AFM) and nonmagnetic (NM) 
states of CaN, where RS structure is assumed for both substances.
}
\end{figure}

We show in Fig. \ref{fig:3} the difference in the total energies for
the ferromagnetic, antiferromagnetic and nonmagnetic states for RS-CaN.  
The energy difference between the ferromagnetic and antiferromagnetic
states for RS-CaN is more than 0.1 eV.
These values are quite large. 

\begin{figure}[tb]
\includegraphics[width=8cm]{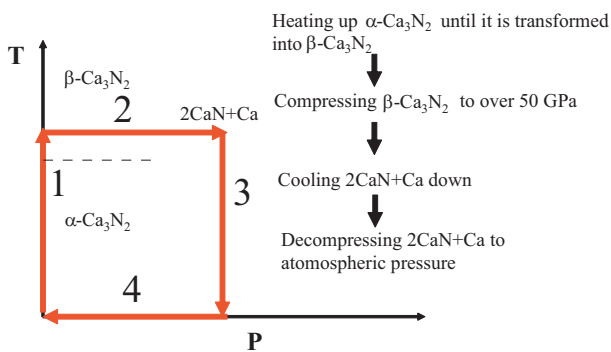}
\caption{\label{fig:4} (Color online)
A promising synthesis process for CaN. 
} 
\end{figure}

We further confirmed the structural stability of CaN having the
RS structure by using first-principles molecular dynamics 
simulations.  
The simulations were performed using the cubic unit cell, which contains
eight atoms and is four times larger than the primitive cell, as the
starting cell.  
We did not assume any symmetry for the atomic positions, that is,
we employed a triclinic structure.
We performed two types of molecular dynamics simulations; (1) a first-principles 
damped dynamics simulation using a variable cell, and (2) a first-principles 
constant-pressure and temperature molecular dynamics simulation using the 
Parrinello-Rahman method, where we used the velocity scaling method 
to control the temperature. 
The purpose of simulation (1) was to confirm whether the RS structure is
at least metastable with respect to changes in the internal atomic
positions and in the shape of the unit cell, while the purpose of 
simulation (2) was to confirm the stability of the RS structure at
finite temperature and atmospheric pressure.  
From simulation (1) we confirmed that atomic forces become negligibly small 
(less than 1.0 $\times$ $10^{-4}$ Ry/a.u.) and that atomic positions in 
the RS structure are at least metastable. 
From simulation (2) we confirmed that the RS structure is stable 
at atmospheric pressure and finite temperatures (130 K, 210 K, and 760 K). 
These results support the possibility that CaN having the RS 
structure would be stable if it can be synthesized.

We estimated the formation energy defined by 
$E_{\rm formation}=E_{\rm tot} - (E_{\rm Ca(Sr)}+ E_{\rm N_2}/2)$ 
for the RS, ZB and CsCl structures.  
$E_{\rm Ca(Sr)}$ is the total energy of the bulk Ca (Sr) per atom at a
atmospheric pressure which we evaluated using the fcc structure. 
$E_{\rm N_2}$ denotes the total energy of an isolated N$_2$ molecule. 
The energy of the molecule is calculated using a large unit cell 
(a=18 \AA, for an fcc cell cotaining one molecule). 
The calculated formation energies are -11.2 eV(RS), -10.7 eV(ZB), and
-10.8 eV(CsCl) for CaN, and -10.8 eV(RS), -10.5 eV(ZB), and -10.5
eV(CsCl) for SrN, respectively. 
These results suggest that CaN and SrN having the three structures
investigated, will not decompose once they have been formed. 
Out of these structures, the formation energy of the RS is lower than
that of the other structures for both CaN and SrN. 

Under normal conditions, Ca and N usually form Ca$_3$N$_2$ in the
$Ia\bar{3}$($\alpha$-Ca$_3$N$_2$) structure. 
This structure is transformed to $\beta$-Ca$_3$N$_2$ whose structure is
$P\bar{3}m1$\cite{Pearson} when the temperature is increased. 
We compared the formation energies of RS-CaN with that of 
Ca$_3$N$_2$ for both $\alpha$- and $\beta$-Ca$_3$N$_2$. 
The formation energy of the $\alpha$ phase is about 2.9 eV lower than
that of 2 CaN + Ca.
The formation energy of the $\beta$ phase, however, is about 2.5 eV
higher than that of 2 CaN + Ca.
This fact implies that if CaN is formed using an appropriate method,
it may be more stable than $\beta$-Ca$_3$N$_2$.

\begin{figure}[tb]
\includegraphics[width=7.5cm]{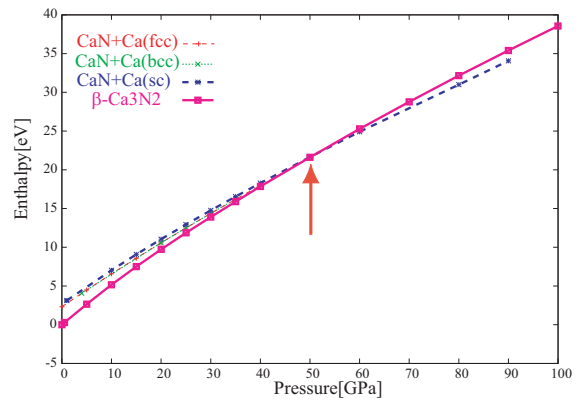}
\caption{\label{fig:5} (Color online)
The enthalpy of $\beta$-Ca$_3$N$_2$ and the sum, 2CaN+Ca, as a 
function of pressure. 
Three types of structure of Ca are assumed and plotted. 
}
\end{figure}

We now proceed to propose ideas for how CaN and SrN can be synthesized 
in the laboratory. 
In this paper, we only consider RS-CaN.
Based on the above results, several possible methods and processes
suggest themselves.  
One of them is high pressure and high temperature synthesis.
We propose the synthesis process based on first-principles calculations.  
A scenario for synthesizing RS-CaN is illustrated 
in Fig. \ref{fig:4}.  
Firstly, heat $\alpha$-Ca$_3$N$_2$ until it is transformed into
$\beta$-Ca$_3$N$_2$ whose structure is $P\bar{3}m1$\cite{Pearson}   
and then compress it using a pressure higher than 50 GPa until the
reaction $\beta$-Ca$_3$N$_2$ $\rightarrow$ 2CaN + Ca occurs. 
After that, cool the product and decompress it to atmospheric pressure. 
In order to evaluate the transition pressure, we give the enthalpy 
differences in Fig. \ref{fig:5}, where we show the enthalpy of  
$\beta$-Ca$_3$N$_2$ and that of the sum of the two phases, 2CaN+Ca. 
We assumed the structures of Ca as fcc, bcc, or sc. 
As the pressure is increased, the structure of Ca is transformed from
the fcc into bcc (at 20 GPa) into sc (at 32 GPa). 
The enthalpy of 2CaN+Ca (sc) becomes lower than that of
$\beta$-Ca$_3$N$_2$ at about 50 GPa.  
Based on the calculations of the formation energy of RS-CaN and
the constant pressure and temperature first-principles molecular
dynamics simulation,  2CaN+Ca(sc) should not decompose once it has been
formed.  
This result strongly suggests that $\beta$-Ca$_3$N$_2$ is transformed 
into RS-CaN with sc Ca.  
The transition temperature from $\alpha$-Ca$_3$N$_2$ to the
$\beta$-Ca$_3$N$_2$ is only represented by a dashed line, since it has 
not been precisely determined experimentally.  
Based on this scheme, laser heating may heat the sample to a
temperature that is too high for the $\beta$-phase.   
Accurate experimental determination of the phase diagram of Ca$_3$N$_2$ 
is required to determine the transition temperatures.

In the view of experiment, we might consider $\beta$-Ca$_3$N$_2$ +N$_2$ 
$\rightarrow$ 6 CaN reaction.
In this process N$_2$ is liquid and is used to compress
hydrostatistically. 
We have not checked the influence of pressure for that process yet, and
we do not deny the possibility.

The other method having a high probability of success is molecular beam
epitaxy. 
To synthesize the RS structure, we propose a synthesis process
using a superlattice structure, similar to the fabrication process of
CrAs/GaAs which used a superlattice structure\cite{Akinaga}. 
On surfaces, processes may occur which are very different from those
that occur in the bulk material and, consequently structures that are
different from those in the bulk may be produced. 
This is the case of ZB-CrAs, which is unstable according to total energy
calculations.   
We do not deny the existence of other synthesis processes, for example,
heating to a liquid state and then compressing to an appropriate
pressure, similar to the process for synthesizing PtN.  
Our suggestions are only two of several possible processes.
We hope that our suggestion inspires experimentalists to synthesize CaN
and SrN.

In conclusion, we have designed and investigated new ferromagnetic
nitrides, CaN and SrN. 
These are half-metallic ferromagnets as a result of the polarization of 
$p$-orbitals of N. 
The process for synthesizing RS-CaN using a high pressure is proposed
based on the results of the first-principles calculations. 
The development of experimental techniques may allow us to synthesize
such new materials.

\begin{acknowledgments}
This work was partially supported by a Grant-in-Aid for Scientific Research  
in Priority Areas ``Development of New Quantum Simulators and Quantum 
Design'' (No. 17064013) of The Ministry of Education, Culture, Sports,
Science, and Technology, Japan, a Grant-in-Aid for Scientific
 Research (15G0213), the 21st Century COE Program supported by
 Japanese Society for the Promotion of Science (JSPS). 
The calculation was partly performed using the computer facilities of ISSP, 
University of Tokyo and the supercomputing resources at the Information
 Synergy Center, Tohoku University.
\end{acknowledgments}

\bibliography{CaN.bib}

\end{document}